\documentclass[letter]{jpsj3arxiv}
\usepackage{txfonts}
\usepackage{color}
\usepackage{amssymb}
\usepackage{bm}

\title{
Pressure-Driven Quantum Criticality and $T/H$ Scaling in the Icosahedral Au-Al-Yb Approximant
}

\author{Shuya~Matsukawa$^1$, Kazuhiko~Deguchi$^1$, Keiichiro~Imura$^1$, 
Tsutomu~Ishimasa$^2$, and Noriaki~K.~Sato$^1\thanks{E-mail: kensho@cc.nagoya-u.ac.jp}$}
\inst{$^1$Department of Physics, Graduate School of Science, Nagoya University, Nagoya 464-8602, Japan\\
$^2$Division of Applied Physics, Graduate School of Engineering, Hokkaido University, Sapporo 060-8628, Japan}

\abst{
We report on ac magnetic susceptibility measurements under pressure of the Au-Al-Yb alloy, a crystalline approximant to the 
icosahedral quasicrystal that shows unconventional quantum criticality.
In describing the susceptibility as $\chi(T)^{-1} - \chi(0)^{-1} \propto T^{\gamma}$,
we find that $\chi(0)^{-1}$ decreases with increasing pressure 
and vanishes to zero at the critical pressure $P_{\rm c} \simeq 2$ GPa,
with $\gamma~ (\simeq 0.5)$ unchanged.
We suggest that this quantum criticality emerges owing to critical valence fluctuations.
Above $P_{\rm c}$, the approximant undergoes a magnetic transition at $T \simeq 100$ mK.
These results are contrasted with the fact that, in the quasicrystal, the quantum criticality is robust against the application of pressure.
The applicability of the so-called $T/H$ scaling to the approximant is also discussed.
} 

\begin{document}
\maketitle

Quantum phase transitions that occur at zero temperature have attracted much interest in the last decades. 
In some cases, unconventional superconductivity emerges in the vicinity of the quantum critical point (QCP) where the second-order transition temperature vanishes to zero.
Experimentally, such a transition can be achieved by tuning an appropriate parameter such as pressure. 
Recent progress in material exploitation has revealed the emergence of a new class of quantum criticality that cannot be understood within the framework of conventional theories.~\cite{Si,Imada,Wa,M}

Such a typical example is found in the Tsai-type Au-Al-Yb icosahedral quasicrystal. 
Here, quasicrystals are metallic alloys that possess a nonperiodic yet long-range-ordered (quasiperiodic) arrangement of the atomic structure. 
As the temperature $T$ is lowered toward zero,
the physical properties diverge like $\chi \propto T^{-0.51}$ and $C_{\rm P}/T \propto -\ln T$ (where $\chi$ is the magnetic susceptibility and $C_{\rm P}$ is the specific heat),~\cite{Deguchi}
suggesting that the system is just at the QCP without tuning.

In contrast, we found no divergence in the Au-Al-Yb approximant crystal, a crystalline phase whose unit cell has atomic decorations (i.e., Tsai-type clusters) that look like the quasicrystal (see Fig.~1 of Ref.~\citen{Deguchi}).
Magnetic susceptibility 
is described by 
$1/\chi - 1/\chi(0) \propto T^{0.51}$ [\citen{Deguchi}], 
where the constant term $\chi(0)^{-1}$
ensures the absence of divergence, indicating that the system is away from the QCP.

The origin of the unusual low-temperature properties of the quasicrystal and the approximant is still controversial.~\cite{M, S,J}
Interestingly, these non-Fermi liquid features are similar to those of the heavy fermion crystals $\beta$-YbAlB$_4$ and YbRh$_2$Si$_2$.~\cite{Y1,Y2,Y3,Y4}
Therefore, further study of the Au-Al-Yb systems is helpful for obtaining a deeper understanding of quantum criticality in strongly correlated electron systems.

\begin{figure}[b]
\begin{center}
\includegraphics[clip,scale=0.7]{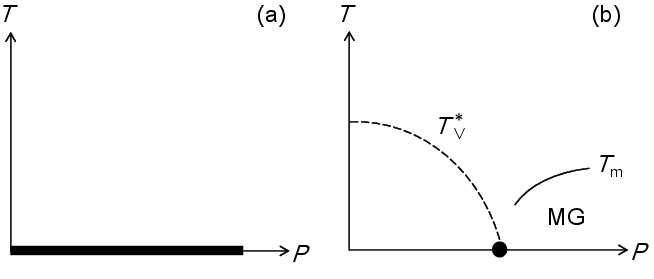}
\end{center}
\caption{
Schematic $P$-$T$ phase diagrams of the Au-Al-Yb quasicrystal (a) and approximant (b). The thick line in (a) indicates ``a collection of QCPs", which may be called the quantum critical line. The closed circle in (b) indicates the QCP. $T_{\rm V}^*$ is a characteristic temperature that is related to valence crossover.~\cite{Wa} $T_m$ and MG indicate a magnetic-ordering temperature and a magnetic ordered state, respectively.
}
\label{fig1}
\end{figure}
The quantum critical behavior of the quasicrystal is robust against the application of hydrostatic pressure ($P$).
The divergence of magnetic susceptibility with the same critical index $\gamma$ was observed even at $P=1.54$ GPa,~\cite{Deguchi} leading us to suggest the $P$-$T$ phase diagram as schematically shown in Fig.~\ref{fig1}(a).
Instead of a quantum critical ``point" in a conventional case, there seems to be a quantum critical ``line" denoted by the thick line. 
This demonstrates that the quasicrystal is distinguished from conventional heavy fermion crystals in which there is a single QCP (or two QCPs as discussed below) and a tuning parameter is needed to control the distance from the QCP.
In contrast, the approximant shows 
a nondiverging feature at ambient pressure;
the susceptibility
saturates at a finite value
$\chi(0)$ with lowering temperature.
However, it is yet unknown how the approximant behaves under pressure.
Here, we study the high-pressure magnetic susceptibility of the approximant crystal
and report the emergence of quantum criticality due to critical valence fluctuations at the critical pressure $P_{\rm c}$, above which the magnetic ordering occurs [see Fig.~\ref{fig1}(b)].
We also study the magnetic field effect on the criticality.
It is known that the susceptibility is suppressed by the dc magnetic field at ambient pressure for both the approximant and the quasicrystal.~\cite{Deguchi} 
Here, we report that the approximant shows a similar suppression of the susceptibility under pressure,
which enables us to examine quantum criticality in terms of $T/H$ scaling relations.

A 1/1 approximant crystal with the nominal composition, Au$_{49}$Al$_{36}$Yb$_{15}$, was prepared by arc-melting the starting materials, namely, 4N (99.99{\%} pure)-Au, 5N-Al, and 3N-Yb, and subsequently annealing the obtained alloy ingot in an evacuated quartz ampoule at 750 Ž for 116 h.~\cite{ref2}
The ac magnetic susceptibility $\chi$ was measured by the conventional mutual inductance method.
A modulation field with a frequency of 100.3 Hz and an amplitude of 0.1 Oe was superimposed on a dc magnetic field
supplied by a superconducting magnet.
For the calibration of the ac magnetic susceptibility, we measured the dc magnetization $M$ using MPMS (Quantum Design) at pressures of up to 1.2 GPa, above which we extrapolated the dc magnetization data. 
Hydrostatic pressure was generated using a NiCrAl-BeCu piston cylinder cell for the measurement of ac susceptibility, together with Daphne oil 7373 as a pressure-transmitting medium. 
The pressure at low temperature was determined from the superconducting transition temperature of indium~\cite{ref3} that was put into the pressure cell together with the sample.

\begin{figure*}[t]
\begin{center}
\includegraphics[scale=0.76]{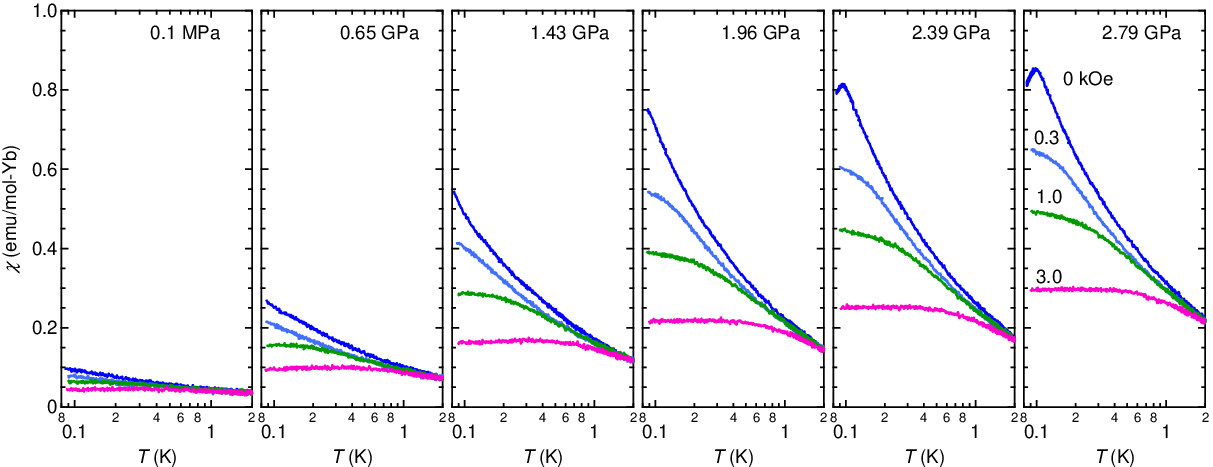}
\end{center}
\caption{
(Color online) Temperature dependence of the ac magnetic susceptibility $\chi$ of the approximant crystal Au$_{49}$Al$_{36}$Yb$_{15}$ in a temperature range of 0.08 $\lesssim T \lesssim$ 2 K at pressures indicated. Applied dc magnetic fields are denoted in the figure of $P=2.79$ GPa.
}
\label{fig2}
\end{figure*}
Figure \ref{fig2} shows the temperature dependence of the ac susceptibility, $\chi(T)$, of the Au-Al-Yb approximant crystal under the dc magnetic field 
$H$ and the hydrostatic pressure $P$. 
Here, $\chi(T)$ is the real part of the ac magnetic susceptibility. 
The low-temperature susceptibility for $H=0$ strongly increases with pressure, whereas the application of the dc field suppresses this increase.
At $P=2.39$ and 2.79 GPa (the highest pressure of the present measurement), a peak structure is formed at $T_{\rm m} \simeq 100$ mK for $H=0$.
It remains unknown
whether this anomaly is associated with an antiferromagnetically long-range ordering or a spin-glass-like short-range ordering.

\begin{figure}[b]
\begin{center}
\includegraphics[scale=0.65]{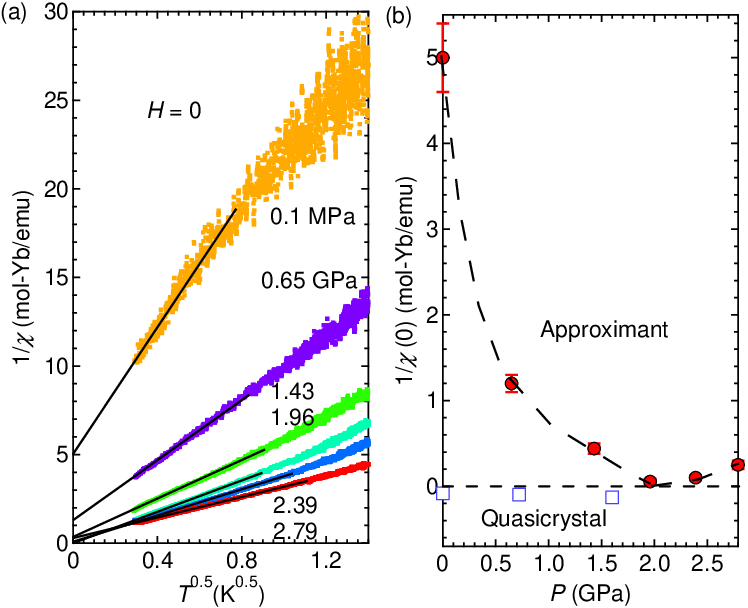}
\end{center}
\caption{
 (Color online) (a) Inverse magnetic susceptibility $1/{\chi}$ vs $T^{0.5}$ of the approximant Au$_{49}$Al$_{36}$Yb$_{15}$ at $H$ = 0 
under pressure ranging from ambient pressure to 2.79 GPa. The straight line indicates the extrapolation to zero temperature to evaluate $1/\chi(0)$.
(b) Pressure dependence of $1/\chi(0)$ of the approximant (circles) and the quasicrystal (squares) at $H$ = 0.
}
\label{fig3}
\end{figure}
On the basis of the results obtained in Ref.~\citen{Deguchi} (see above), we assume the following 
modified Curie--Weiss relation 
for $H = 0$ except in a temperature region near $T_{\rm m}$;
\begin{equation}
{\chi(T)}^{-1}
=T^{0.5}/C + {\chi(0)}^{-1}.
\label{equ1}
\end{equation}
Here, the traditional Curie--Weiss plot where 1/$\chi$ is a straight line as a function of $T$ is replaced by a plot where 1/$\chi$ is a straight line as a function of $T^{0.5}$ with the slope $1/C$ (where $C$ is the Curie constant) and the vertical-axis intercept ${\chi(0)}^{-1}$.
This straight-line feature is confirmed in Fig.~\ref{fig3}(a) over a temperature range between about 85 mK (the base temperature of the experiment) and about 800 mK. 
(When evaluating the exponent $\gamma$ 
 as a free parameter, we 
obtain $\gamma = 0.50 \pm 0.05$ at ambient pressure and $\gamma = 0.50 \pm 0.01$ at $P = 1.96$ GPa.)
Although the exponent $\gamma$
does not depend on pressure, the intercept 
${\chi(0)}^{-1}$ (that measures the distance from the QCP)
approaches zero with increasing pressure.
The latter observation is more clearly seen in Fig.~\ref{fig3}(b): 
$1/\chi(0)$ steeply decreases with increasing pressure and vanishes at around 2 GPa, suggesting the emergence of the QCP there.
This distinguishes the approximant from the quasicrystal in which $1/\chi(0) \sim 0$, independent of pressure, as indicated by the open square.~\cite{Deguchi}

It is now evident that the approximant is qualitatively different from the quasicrystal in the response to hydrostatic pressure,
as schematically illustrated in Fig.~\ref{fig1}.
It seems reasonable to assume that this difference between the quasicrystal and the approximant is ascribed to the absence/presence of the periodicity. 
This is one of the primary results of this study.

\begin{figure}[b]
\begin{center}
\includegraphics[scale=0.68]{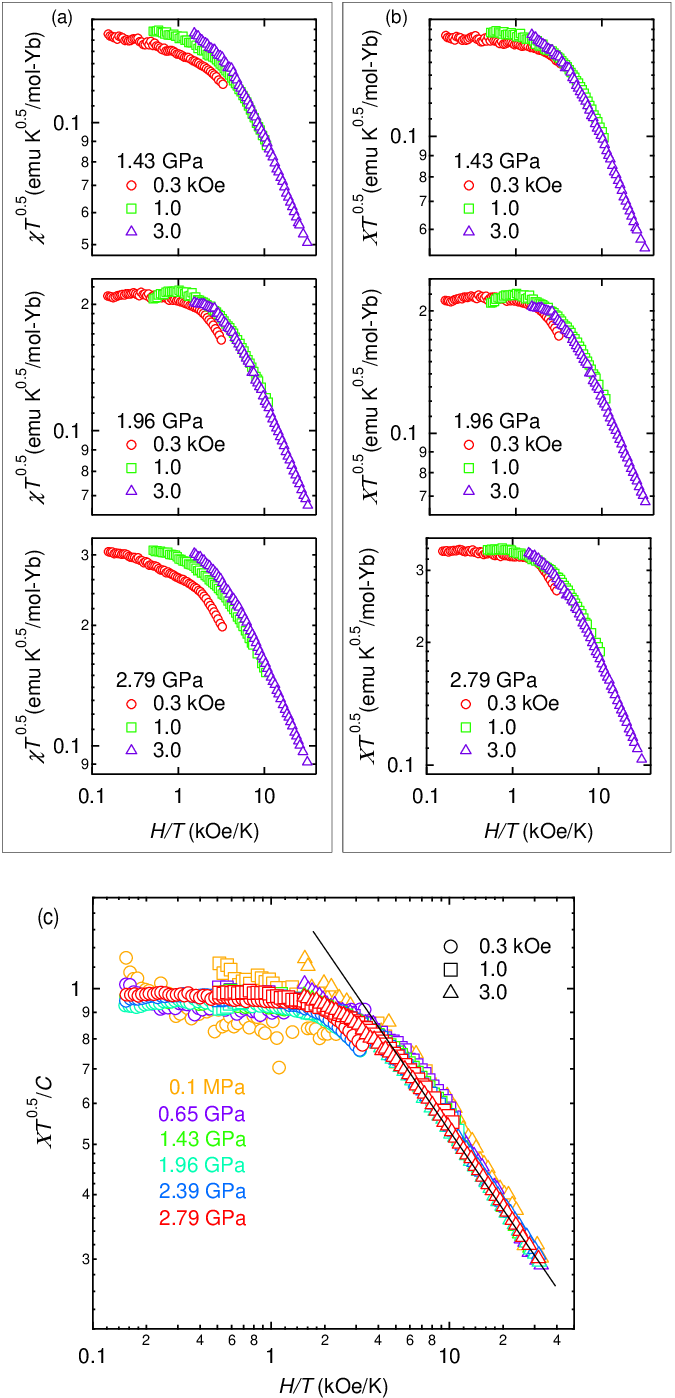}
\end{center}
\caption{
(Color online)
$H/T$ scaling of the approximant Au$_{49}$Al$_{36}$Yb$_{15}$. 
The vertical axis denotes (a) $\chi T^{0.5}$, (b) $XT^{0.5}$, and (c) $XT^{0.5}/C$.
(a) 
The scaling behavior observed at $P=1.96$ GPa ($\simeq P_{\rm c}$) becomes less evident as 
$P$ moves away from $P_{\rm c}$ in either direction.
(b) The critical component $X$ satisfies the scaling relation even for $P \neq P_{\rm c}$. (c) All the data 
fall on a single curve.
The straight line indicates the Fermi liquid limit of the scaling function $\eta^{-1}$. 
}
\label{fig4}
\end{figure}
In what follows, we study 
the magnetic field effect on the criticality. First, we examine
the scaling properties 
of the ac susceptibility $\chi= \frac{d M}{d H}$.
Let us start from the following uniform and static susceptibility that is deduced from the generalized susceptibility proposed for CeCu$_{6-x}$Au$_x$ by Schr\"oder {\it et al.},~\cite{Schroder}
\begin{eqnarray}
\chi^{-1}(H, T) 
&=& \frac{1}{C} \left[ \frac{1}{k_{\rm B}^\alpha} \left( (k_{\rm B}T)^2 + (g\mu_{\rm B}H)^2 \right)^{\alpha/2} + \theta^\alpha \right] \nonumber \\
&=& \frac{1}{C} \eta \left(\frac{H}{T} \right) T^{\alpha} + \frac{1}{C}\theta^{\alpha},
\label{eq:scl0}
\end{eqnarray}
where $k_{\rm B}$, $g$, $\mu_{\rm B}$, and $\theta$ are the Boltzmann constant, effective $g$-factor, Bohr magneton, and Weiss temperature, respectively. 
$\eta$ is a scaling function of the ratio $H/T$ only 
and equal to 1 for  $H/T=0$. The power index $\alpha$ was found to be 0.75 for CeCu$_{6-x}$Au$_x$.
For the present system, we obtain $\alpha=0.5$ by comparing Eq.~(\ref{eq:scl0}) with the phenomenologically determined Eq.~(\ref{equ1}).
(Note that $\eta =1$ for $H=0$.)
Defining $X$ as
\begin{equation}
X^{-1} = \chi^{-1} - \frac{\theta^{0.5}}{C},
 \label{eq:X}
\end{equation} 
we obtain the scaling relation
\begin{equation}
XT^{0.5} = C \eta^{-1}.
\label{eq:scl}
\end{equation}
At the QCP where $\chi(0)^{-1}=\theta^{0.5}/C=0$, we have $X=\chi$ and hence the scaling relation
\begin{equation}
\chi T^{0.5} = C \eta^{-1}, \mbox{ at QCP}.
\label{eq:scl2}
\end{equation}
We plot $\chi T^{0.5}$ and $XT^{0.5}$
as functions of $H/T$ in Figs.~\ref{fig4}(a) and \ref{fig4}(b), 
respectively.
In Fig.~\ref{fig4}(a), we find that 
only the data of $P=1.96$ GPa close to $P_{\rm c}$ collapse on a single curve,
meaning that the scaling relation Eq.~(\ref{eq:scl2}) holds in the present system.
Equation (\ref{eq:scl}) also holds because the data even at $P \neq P_{\rm c}$ fall on a single curve, as seen in Fig.~\ref{fig4}(b).
Taking into account the fact that $C$ depends on pressure [see Fig.~\ref{fig3}(a)],
we plot $XT^{0.5}/C$ in Fig.~\ref{fig4}(c);
although the ambient-pressure data scatter owing to their small susceptibility,
all pressure data taken here fall on a single curve, 
which manifests that $X T^{0.5}/C$ is a function of $H/T$ only.

Equation (\ref{eq:scl}) is confirmed more quantitatively. According to
the asymptotic form of the scaling function $\eta^{-1}$,
\begin{equation} 
\eta^{-1} 
\left( \frac { H }{ T }  \right) 
\sim \left\{ 
\begin{array}{ll}
1-\frac {1}{4} \left( \frac { g \mu_{\rm B} H }{k_{\rm B}T } \right)^2, & \mbox{ for 
$ T \gg H$, 
} \\
\left( \frac { g\mu_{\rm B } H }{ k_{ \rm B }T } \right)^{ -1/2 }, & \mbox{ for 
$T \ll H$.
}
\end{array}
\right.  
\label{equ5}
\end{equation}
the data shown in Fig.~\ref{fig4}(c) saturate at unity
in the non-Fermi liquid limit ($H/T \ll 1$), 
while in the opposite limit of Fermi liquid ($H/T \gg 1$), 
they decrease linearly with increasing $H/T$.
This indicates that the magnetic field drives the system to the Fermi liquid regime.

Next, we discuss that
the unusual exponent $\alpha = 0.5$ obtained above can be deduced from
the following low-temperature free energy, which was proposed for the analysis of the quantum criticality of $\beta$-YbAlB$_4$ by Matsumoto {\it et al.},~\cite{Y2}
\begin{equation} 
F=-\frac {1}{ (k _{\rm B} \tilde{T})^{ 1/2 } } { \left[ \left( g\mu_{\rm B}H \right)^2 + \left( k_{\rm B}T \right) ^{ 2 } \right]  }^{ 3/4 },
\label{equ2}
\end{equation}
where $\tilde{T}$ is a characteristic temperature.
By differentiating $F$ with respect to $H$, we obtain the ac susceptibility 
and the dc susceptibility $M/H$ (where $M$ is a magnetization)
as
\begin{eqnarray} 
\chi(T,H) T^{1/2}
&=& 
\psi \left(\frac{H}{T}\right), ~~~~~~{\rm for}~ H\geq 0,
\label{equ4} \\
\frac{M}{H} H^{1/2}
&=&  
\varphi \left(\frac{T}{H}\right), ~~~~~~{\rm for}~ H>0,
\label{equ4a} 
\end{eqnarray}
where $\psi $ and $\varphi $ are scaling functions, and $\psi = \frac{3}{2} \frac{\left(g{\mu}_{\rm B}\right)^2}{k_{{\rm B}}\tilde{T}^{1/2}}$ for $H=0$.
Equation (\ref{equ4}) is equivalent to Eq.~(\ref{eq:scl2}) when $C\eta^{-1} = \psi$, and a comparison of Eq.~(\ref{equ4}) with Eq.~(\ref{eq:scl0}) yields $\alpha = 0.5$.

\begin{figure}[t]
\begin{center}
\includegraphics[scale=0.61]{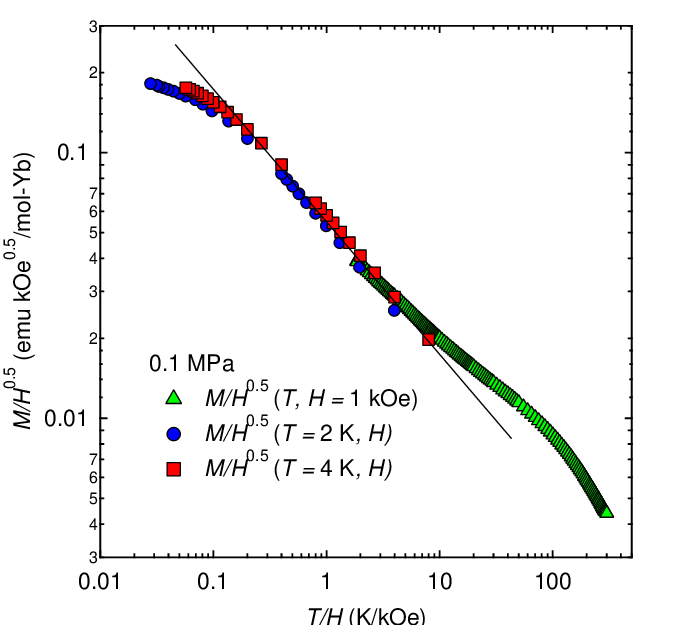}
\end{center}
\caption{
(Color online) $T/H$ scaling of dc susceptibility, $(M/H)H^{0.5}$, in a temperature interval between 2 K and room temperature. 
Note that the horizontal axis denotes $T/H$, instead of $H/T$ in Fig.~\ref{fig4}.
Triangles indicate a temperature-sweep experiment, and circles and squares indicate a dc-field-sweep experiment at $T=$ 2 and 4 K, respectively. 
}
\label{fig5}
\end{figure}

Then, we test the scaling relation Eq.~(\ref{equ4a}) for the dc magnetization
using data taken at high temperatures (2 K $\lesssim T \lesssim$ room temperature) and high dc fields ($ H \lesssim 70$ kOe) (see Fig.~\ref{fig5}).
(We used the ambient-pressure data instead of the high-pressure one
because the high-pressure magnetization contains a non-negligible contribution from the pressure cell and it was difficult to accurately separate 
the sample magnetization from the total magnetization.)
For $T/H > 10$ (i.e., $T > 10 $ K for $H= 1$ kOe), 
we find a deviation from the expected behavior (denoted by the thin line) from the asymptotic form
\begin{equation} 
\varphi \left( \frac { T }{ H }  \right) 
\sim 
\frac{3(g \mu_{\rm B})^{3/2}}{2\left( k_{{\rm B}}\tilde{T}\right)^{1/2}}
\left( \frac{k_{\rm B} T}{g\mu_{\rm B} H} \right)^{-1/2}, 
\mbox{ for $ T \gg H$}.
\label{equff}
\end{equation}
This is ascribed to the crystal field effect 
because the 4$f$ electrons are localized at high temperatures as evidenced by
the conventional Curie--Weiss behavior in $\chi(T)$.~\cite{Deguchi,matsukawa}

It is now clear that 
the $T/H$ scaling holds (except in the very high temperature region) in the approximant like CeCu$_{6-x}$Au$_x$ and $\beta$-YbAlB$_4$.
We note that the same scaling is also applicable to the Au-Al-Yb quasicrystal (not shown here).
In all four systems, therefore,
the critical field $H_{\rm c}$ of the quantum phase transition is zero
because a finite $H_{\rm c}$ would require that the argument of the scaling functions is the ratio $T/|H - H_{\rm c}|$.~\cite{Y2}
This 
may lead to a similar $H$-$T$ phase diagram for those systems.
In contrast,
the $P$-$T$ phase diagram strongly depends on the system as demonstrated in Fig.~\ref{fig1}; for CeCu$_{5.8}$Au$_{0.2}$ and $\beta$-YbAlB$_4$, see Refs.~\citen{St} and \citen{Tomita}, respectively.

Let us discuss a possible origin of the QCP.
As presented above, the uniform susceptibility shows the power law divergence with the critical index $\gamma \simeq 0.5$ at $P=P_{\rm c}$.
This divergence cannot be understood from the conventional
magnetic QCP because the high-pressure magnetic state is not ferromagnetic but antiferromagnetic (or spin-glass-like).
Instead, the divergence with the above critical index 
can be explained
by the critical valence fluctuation model proposed by Watanabe and Miyake.~\cite{Watanabe}
Remembering that the valence in the present system fluctuates between Yb$^{2+}$ and Yb$^{3+}$ as evidenced by X-ray absorption experiments \cite{Watanuki} and photoemission experiments,~\cite{Matsunami} we suggest that the QCP arises from the critical valence fluctuations.
This interpretation is supported by the theoretical suggestion that the scaling relation Eq.~(\ref{equ4a}) is deduced from the critical valence fluctuation model.~\cite{Watanabe1}

Finally, we comment on the $P$-$T$ phase diagram of the approximant.
In Fig.~\ref{fig1}(b), we plot the valence-fluctuation QCP.
We found the emergence of the magnetic (short-range or long-range) ordering state at pressures above $P_{\rm c}$.
Therefore, it is possible that there is another QCP of magnetic origin at which $T_{\rm m}$ vanishes.
The interplay between magnetic and valence instabilities was predicted using the critical valence fluctuation model,~\cite{Wa} and was  observed experimentally in YbNi$_3$Ga$_9$.~\cite{Matsubayashi}
To compare these 
results with the present system, a detailed $P$-$T$ phase diagram of the approximant should be constructed.

In summary, we measured the ac magnetic susceptibility of the approximant crystal Au$_{49}$Al$_{36}$Yb$_{15}$.
As $T \rightarrow 0$, the uniform susceptibility diverges at the critical pressure $P_{\rm c} \simeq 2$ GPa, which we ascribed to the critical valence fluctuations.
At pressures above $P_{\rm c}$, the magnetic transition occurs at $T_{\rm m} \simeq 100$ mK,
although it remains to be revealed in the future whether this transition is of long- or short-range nature. 
This pressure effect is remarkably contrasted with the robustness of the quantum criticality in the quasicrystal against the application of pressure,
which we ascribe to the difference in the presence/absence of the periodicity.
For the magnetic field effect, on the other hand, the approximant satisfies the $T/H$ scaling applied to CeCu$_{6-x}$Au$_x$ and $\beta$-YbAlB$_4$, meaning that there is no critical field $H_{\rm c}$.
In conclusion, we found that the pressure and magnetic field play different roles in quantum criticality.

The authors thank Yukinori Tanaka and Shin Yamamoto for support of the experiments, and Kazumasa Miyake for careful reading of the manuscript. This work was financially supported by a Grant-in-Aid for Scientific Research from JSPS, KAKENHI (Nos. 
26610100, 15H02111, and 15H03685) and Program for Leading Graduate Schools ``Integrative Graduate Education and Research Program in Green Natural Sciences", MEXT, Japan, and also 
by Toyota Physical \& Chemical Research Institute.


\begin{thebibliography}{9}
\bibitem{Si} Q.~Si, S.~Rabello, K.~Ingersent, and J.~L.~Smith, Nature {\bf 413}, 804 (2001).
\bibitem{Imada} M.~Imada, T.~Misawa, and Y.~Yamaji, J. Phys.: Condens. Matter {\bf 22}, 164206 (2010).
\bibitem{Wa} S.~Watanabe and K.~Miyake, J. Phys.: Condens. Matter {\bf 23}, 094217 (2011).
\bibitem{M} S.~Watanabe and K.~Miyake, J. Phys. Soc. Jpn. {\bf 82}, 083704 (2013).
\bibitem{Deguchi} K.~Deguchi, S.~Matsukawa, N.~K.~Sato, T.~Hattori, K.~Ishida, H.~Takakura, and T.~Ishimasa, Nat. Mater. {\bf 11}, 1013 (2012). 
\bibitem{S} V.~R.~Shaginyan, A.~Z.~Msezane, K.~G.~Popov, G.~S.~Japaridze, and V.~A.~Khodel, Phys. Rev. B {\bf 87}, 245122 (2013).
\bibitem{J} E.~C.~Andrade, A.~Jagannathan, E.~Miranda, M.~Vojta, and V.~Dobrosavljevic, Phys. Rev. Lett.
{\bf 115}, 036403 (2015).
\bibitem{Y1} S.~Nakatsuji, K.~Kuga, Y.~Machida, T.~Tayama, T.~Sakakibara, Y.~Karaki, H.~Ishimoto, S.~Yonezawa, Y.~Maeno, E.~Pearson, G.~G.~Lonzarich, L.~Balicas, H.~Lee, and Z.~Fisk,
Nat. Phys. {\bf 4}, 603 (2008).
\bibitem{Y2} Y.~Matsumoto, S.~Nakatsuji, K.~Kuga, Y.~Karaki, N.~Horie, Y.~Shimura, T.~Sakakibara, A.~H.~Nevidomskyy, and P.~Coleman,
Science {\bf 331}, 316 (2011).
\bibitem{Y3} O.~Trovarelli, C.~Geibel, S.~Mederle, C.~Langhammer, F.~M.~Grosche, P.~Gegenwart, M.~Lang, G.~Sparn, and F.~Steglich,
Phys. Rev. Lett. {\bf 85}, 626 (2000).
\bibitem{Y4} P.~Gegenwart, T.~Westerkamp, C.~Krellner, Y.~Tokiwa, S.~Paschen, C.~Geibel, F.~Steglich, E.~Abrahams, and Q.~Si,
Science {\bf 16}, 969 (2007).
\bibitem{ref2} T.~Ishimasa, Y.~Tanaka, and S.~Kashimoto, Phil. Mag. {\bf 91}, 4218 (2011).
\bibitem{ref3} L.~D.~Jennings and C.~A.~Swenson, Phys. Rev. {\bf 112}, 31 (1958).
\bibitem{Schroder} A.~Schr\"{o}der, G.~Aeppli, R.~Coldea, M.~Adams, O.~Stockert, H.~v.~L\"{o}hneysen, E.~Bucher, R.~Ramazashvili, and P.~Coleman, Nature {\bf 407}, 351 (2000).
\bibitem{matsukawa} S.~Matsukawa, K.~Tanaka, M.~Nakayama, K.~Deguchi, K.~Imura, H.~Takakura, S.~Kashimoto, T.~Ishimasa, and N.~K.~Sato, J. Phys. Soc. Jpn. {\bf 83}, 034705 (2014).
\bibitem{St} O.~Stockert, F.~Huster, A.~Neubert, C.~Pfleiderer, T.~Pietrus, B.~Will, and H.~v.~L\"{o}hneysen, Physica B {\bf 312}-{\bf 313}, 458 (2002).
\bibitem{Tomita} T.~Tomita, K.~Kuga, Y.~Uwatoko, P.~Coleman, and S.~Nakatsuji, Science {\bf 349}, 506 (2015).
\bibitem{Watanabe} S.~Watanabe and K.~Miyake, Phys. Rev. Lett. {\bf 105}, 186403 (2010).
\bibitem{Watanuki} T.~Watanuki, S.~Kashimoto, D.~Kawana, T.~Yamazaki, A.~Machida, Y.~Tanaka, and T.~J.~Sato, Phys. Rev. B {\bf 86}, 094201 (2012).
\bibitem{Matsunami} 
M. Matsunami, A. Chainani, M. Taguchi, M. Oura, S. Shin, T. Hajiri, S. Kimura, T. Takeuchi, K. Tamasaku, Y. Tanaka, T. Ishikawa, T. Ebihara, K. Imura, K. Deguchi, N. K. Sato, and T. Ishimasa, unpublished. 
\bibitem{Watanabe1} S.~Watanabe and K.~Miyake, J. Phys. Soc. Jpn. {\bf 83}, 103708 (2014).
\bibitem{Matsubayashi} K.~Matsubayashi, T. Hirayama, T. Yamashita, S. Ohara, N. Kawamura, M. Mizumaki, N. Ishimatsu,
S. Watanabe, K. Kitagawa, and Y. Uwatoko, Phys. Rev. Lett. {\bf 114}, 086401 (2015).



\end{thebibliography}
\end{document}